\documentclass[final,3p,times, two column]{elsarticle}
\usepackage{graphicx}
\usepackage{amsmath}
\usepackage{amssymb}
\usepackage{hyperref}

\begin{document}

\title{Physical properties in nano-crystalline Ho$_2$CoMnO$_6$}

\author{Ilyas Noor Bhatti\corref{cor2}}\address{Department of Physics, Jamia Millia Islamia University, New Delhi - 110025, India.}
\ead{inoorbhatti@gmail.com}
\author{Imtiaz Noor Bhatti}\address{Department of School Education, Government of Jammu and Kashmir, India.}
\author{Rabindra Nath Mahato}\address{School of Physical Sciences, Jawaharlal Nehru University, New Delhi - 110067, India.}
\author{M. A. H. Ahsan}\address{Department of Physics, Jamia Millia Islamia University, New Delhi - 110025, India.}

\begin{abstract}
3$d$ based double perovskite materials have received much attention in recent years due to their exotic magnetic structure and magneto-electric coupling. In this work we have prepared and studied the nano-crystalline sample of Ho$_2$CoMnO$_6$. Structural, magnetic, Raman and dielectric properties have been studied in detail. The structural analysis shows that the sample crystallize in monoclanic crystal structure with $\textit{P2$_1$/n}$ phase group. The X-ray photoelectron spectroscopy have been employed to confirms the charge state of  cations presents in the material.  Magnetic study shows that the sample undergoes a paramagnetic to ferromagnetic phase transition around $T_c$ $\sim$85 K. The isothermal magnetization measurements shows hysteresis curve hence confirm ferromagnetic behavior at low temperature. Temperature dependent Raman study reveals that there is spin phonon coupling in the sample marked by deviation in phonon mode from anharmonic behavior. Dielectric response of Ho$_2$CoMnO$_6$ shows the large dispersion and large dielectric constant. Impedance spectroscopy and electrical modulus study reveal that system shows deviation from ideal Debye model. AC conductivity have been studied as a function of both temperature and frequency. We found that the conduction mechanism is obeyed by Jonscher's model. The exponent factor $n$ is suggest that the material deviates from ideal Debye model.
\end{abstract}

\maketitle
\section{Introduction}
Magnatoelectric and multiferroic materials specially perovskite and double perovskites oxides with rare earth and transition  metals have been the focus of study due to various exotic phenomena like magnatoelectric/magnatodielectric effects, tunable pyroelectric and ferroelectric properties.\cite{ere, che, sug, van} These materials shows low temperature ferroelectic transition and have spontaneous polarization and also shows strong magnetic effects, where dielectric properties are controlled with magnetic field. Several material have been investigated to achieve the desired interplay of electric and magnetic properties for-instance, (Gd/Tb/Dy)MnO$_3$/HoMnO$_3$/YMnO$_3$.\cite{kim1, kim2, goto, yama, ishi, lor} The multiferric perovskite magnates HoMnO$_3$ with ferroelectic transition and large polarization $\sim$ $56 mC/m^2$ shows giant magnetocaloic effects and was reported by Hur et al.\cite{hur} Further, multeroferroic properties have been observed in A-type antiferromagnet Sm$_x$Y$_1$-xMn$O_3$ .\cite{flynn} Exchange intersection ferrielectric in multiferric HoMn$O_3$ have been reported by Lee et al. \cite{lee} 

Recently 3d based double perovskite R$_2$CoMnO$_6$ based R=rare earth elements have been receiving great attraction of researchers due to their electric and magnetic properties. In theses compounds a $Co^{2+}$ with electronic configuration $d^7$ t$_2g^5$ e$_g^2$ and Mn$^4+$ with $d^3$ t$_2g^3$ e$_g^0$ ions are presents. The super exchanges interactions between $Co^{2+}$ and $Mn^4+$ give rise to ferromagnetic ordering in most of these materials.\cite{goodenough} The magnates double peroskites with heavier rare earth elements shows remarkable properties like magnetization reversal and inverse exchange based in Er$_2$CoMnO$_6$,\cite{banerjee} anisotropy magnetic properties and giant magnatocaloric effect in Tb$_2$CoMnO$_6$,\cite{moon} negative magnatocapaictance Yb$_2$CoMnO$_6$ have also been observed. However Ho$_2$CoMnO$_6$ is less studied. Ho having high magnetic moment and relatively small size compared to other rare-earth would be quite interesting to investigate. In this paper we investigation the structural magnetic and dielectric properties of Ho$_2$CoMnO$_6$.

In this paper we have investigated structural, magnetic and dielectric properties of the nano-crystalline Ho$_2$CoMnO$_6$. Structural investigation show that the samples crystallize in monoclinic structure and adopt \textit{P2$_1$/n} space group. Charge state of various elements present in the material is confirmed using X-ray photoelectron spectroscopy. Magnetization study shows that the material is ferromagnetic in nature and shows phase transition around 85 K. The effective magnetic moment obtained experimentally as 5.25 $\mu_B$/f.u. is found close to calculated value. Raman study shows that spin phonon coupling is present as indicated by mode frequency deviation from anharmonic behavior across $T_c$. Dielectric measurements have been performed on Ho$_2$CoMnO$_6$ with both temperature and frequency dependence. We found a very high dielectric constant which increases with decreasing frequency. The tangent loss shows the relaxation mechanism is active and it follows thermally activated behavior. The impedance spectroscopy reveals that relaxation time is distributive and shows deviation from ideal Dabye's model. AC conductivity have been studied as a function of both temperature and frequency. We found that the conduction mechanism is obeyed by Jonscher's model. The exponent factor $n$ is suggestive that the material deviates from ideal Debye's model.

\section{Experimental details}
The nano-crystalline sample Ho$_2$CoMnO$_6$ was prepared by Sol gel method. We have taken the ingredients, Ho$_2$O$_3$, Co(NO$_3$)$_2$.6H$_2$O and C$_4$H$_6$MnO$_4$.4H$_2$O in stoichiometric ratio.  First we dissolved these ingredients in water in separate beakers with continuous stirring, where Co(NO$_3$)$_2$.6H$_2$O and C$_4$H$_6$MnO$_4$.4H$_2$O were completely dissolved in water and make clear solutions. However, Ho$_2$O$_3$ did not dissolve in water, so we added HNO$_3$ in the mixture drop by drop stirred on magnetic stirrer at 70 $^o$C continuously. After 45 minute the solution become clear and Ho$_2$O$_3$ was competely dissolved. These three solution were added in 400 ml beaker with water and placed on the magnetic stirrer for 24 hours for reaction to take place. The solution was turned into the gel form due to evaporation of water. At that time the temperature is increase to 200 $^o$C and the gel is started to dry, after complete dry out of the remain in the beaker was collected and ground in mortar and pestles for 25 minutes to make fine powder. The obtained powder is then placed in furnace for oxidation at 900$^o$C.  The phase purity of prepared sample is checked by powder X-ray differection (XRD) study. XRD data collected at room temperature in 2$\theta$ range 10$^o$ -90$^o$ with step size of 0.02$^o$ and scan rate of 2$^o$/min. Crystal structural analysis was done by  Rietveld refinement of XRD data using Fullprof program. The XPS measurements were performed with base pressure in the range of $10^{−10}$ mbar using a commercial electron energy analyzer (Omnicron nanotechnology) and a non-monochromatic Al$_{K\alpha}$ X-ray source (h$\nu$ = 1486.6 eV). The XPSpeakfit software has been used to analyze the XPS data. The samples used for XPS study are in pellet form where an ion beam sputtering has been done on the samples to expose clean surface before measurements. Magnetic data was collected by Physical Properties Measurement System by Cryogenic Inc. Temperature dependent Raman spectra have been recorded using Diode based laser ($\lambda$ = 473 nm) coupled with a Labram-HR800 micro-Raman spectrometer. It is a single spectrometer with 1800 grooves/mm grating and a peltier cooled CCD detector with a overall spectral resolution of $\sim$ 1 cm$^{−1}$. For the low temperature Raman measurements, the material has been mounted on a THMS600 stage from Linkam UK, with temperature stability of $\pm$0.1 K. Temperature as well as frequency dependent dielectric measurements have been performed on home made  dielectric setup in the frequency range from 1 Hz to 5.5 MHz. The measurements were carried out in temperature range 20 K to 300 K the low temperature is achieved by close cycle refrigerator 

\begin{figure}[t]
	\centering
		\includegraphics[width=8cm]{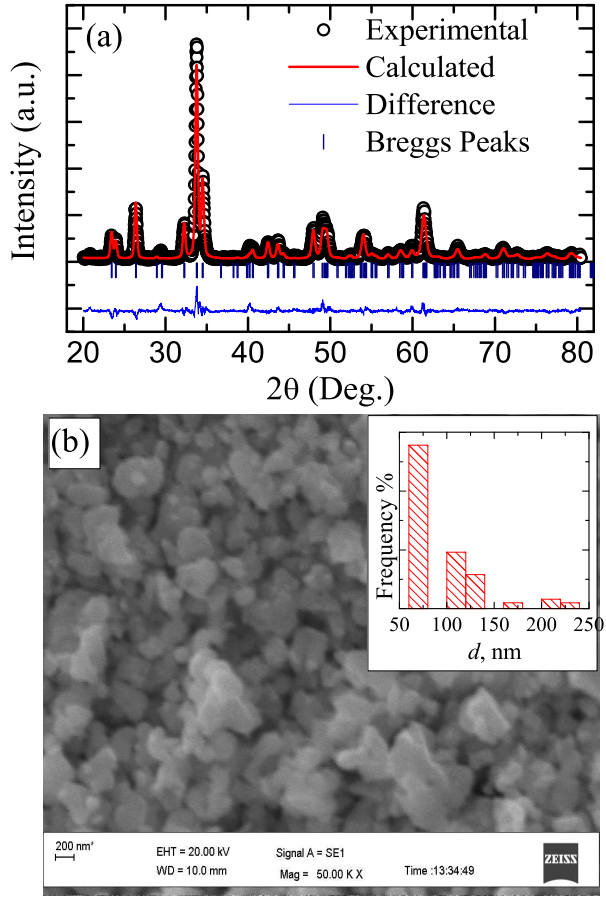}
\caption{(Color online) X-ray diffraction pattern along with Rietveld refinement for nano-crystalline Ho$_2$CoMnO$_6$ (b) Representative SEM image use to estimate particle size. Inset shows grain size distribution.}
	\label{fig:Fig1}
\end{figure}

\section{Result and Discussions}
\subsection{Structural study}
Sample in powder form is characterization by X-ray diffraction (XRD) at room temperature for structural study. The powder diffraction data is analyzed with Rietveld refinement by using Fullprof software. Fig. 1a) represent the XRD pattern along with rietveld refinement for Ho$_2$CoMnO$_6$. Black open circle in the diagram represents original experiment data, red solid lines are represents calculated pattern, with blue thin lines are difference and  navy blue bars represents Braggs positions. Rietveld refinement of the XRD data was fine and sample is in single phase. The analysis of XRD data with rietveld refinement shows that the sample crystallize in monoclinic crystal structure and adopt \textit{P2$_1$/n} space group.  The fitting parameters χ 2 and R$_{wp}$/R$_{exp}$ ratio which gives the goodness of fit of experimental data with the model are found to be 2.05 and 1.27 and are quite acceptable.\cite{ilyas1, bhatti1, bhatti2} The lattice parameter  \textit{a},\textit{b} and \textit{c} at room temperature are obtained as  5.1990(2) $\AA$, 5.5550(8) $\AA$ and 7.4289(5) $\AA$ respectively for Ho$_2$CoMnO$_6$. The unit cell volume and $\beta$ values for Ho$_2$CoMnO$_6$ are found to be 214.5499 $\AA^3$ and 90.037$^o$ respectively. Further, to obtan the partical size in the nano-crystaline sample of Ho$_2$CoMnO$_6$ we have performed the scanning electron microscopy. Fig. 1(b) shows the SEM image obtained for the nano-crystalline structure. The SEM image is analyzed using ImageJ software. We observed that the average grain size for present compound is $\sim$76.65 nm. Further, we have potted the grain size distribution as shown in inset of Fig. 1b.  

\begin{figure}[th]
	\centering
		\includegraphics[width=8cm, height=12cm]{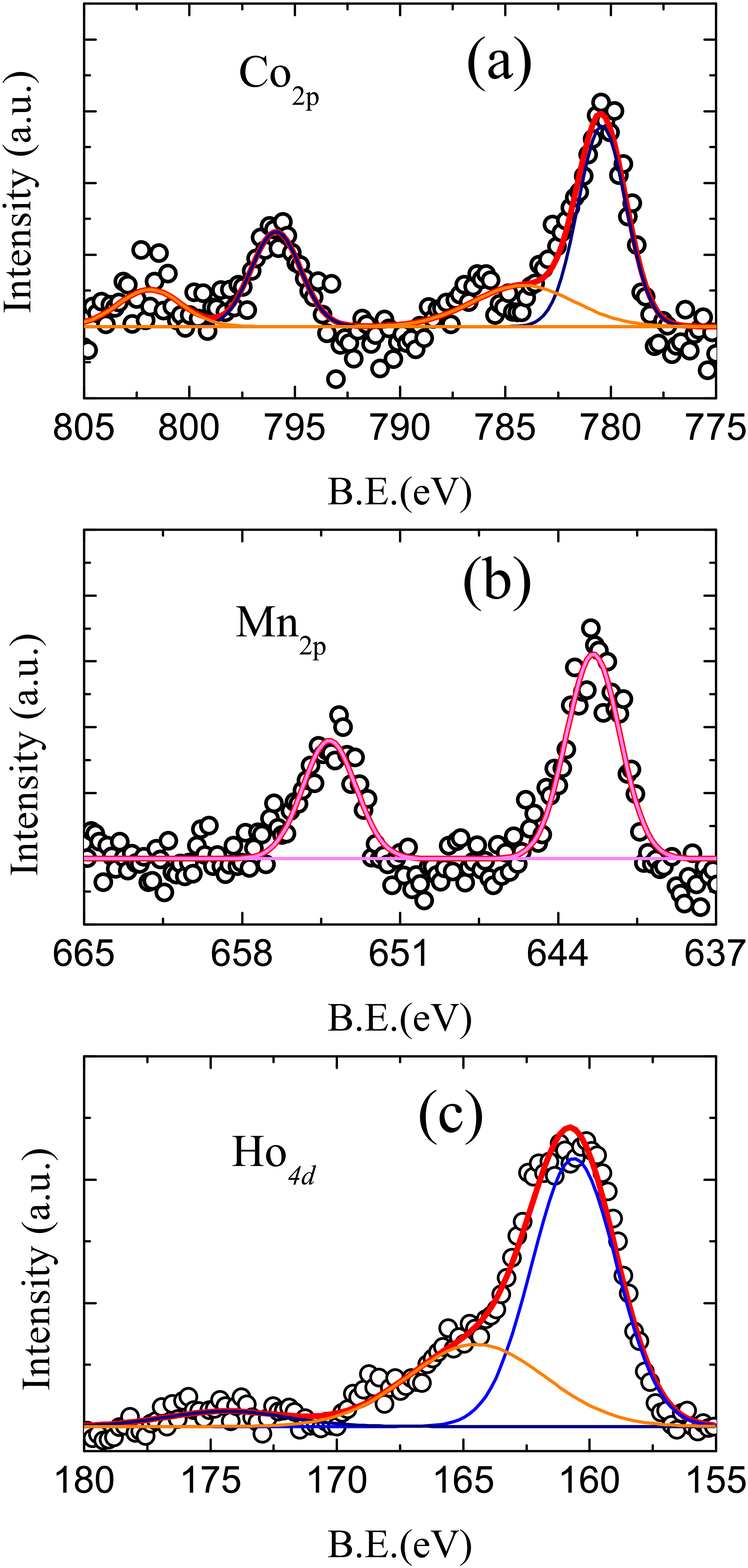}
\caption{(Color online) (a) The XPS core level spectra of Co 2$p$ (b) XPS core level spectra of Mn 2$p$. (c) XPS core level spectra of Ho 4$d$ for Ho$_2$CoMnO$_6$. In the figure the red solid line is the overall envelop of the XPS spectrum and the other colored solid lines are the other respective fitted peaks.}
	\label{fig:Fig2}
\end{figure}

\subsection{X-ray photo-electron spectroscopy (XPS)}
Understanding the oxidation state of elements present in the material gives vital information about the physical properties of materials and underlying physics. Here we have employed XPS to understand the charge state at surface of Ho$_2$CoMnO$_6$. We have employed the XPS to study the cationic charge state of Co, Mn and Ho in Ho$_2$CoMnO$_6$. XPS spectrum of Co 2$p$ is shown in Fig. 2a in which the open black circle are the experimental data the red line is overall envelop of the spectrum the solid blue lines are the Co 2$p$ peaks where as orange solid lines are the satellite peaks. It is evident from the Fig. 2a there are two peaks located at 780.4 eV and 795.8 eV for Co 2$p$$_{3/2}$ and Co 2$p$$_{1/2}$  respectively resulted from spin orbital splitting of 2$p$ orbital with splitting energy of 15.4 eV. Beside the Co 2$p$ peaks two satellite peaks have also been observed close to Co 2$p$ peaks are in agreement with literature.\cite{qiu} The peaks locations of Co 2$p$ core level indicates that the Co cations presents in +2 oxidation states.

The measured XPS spectrum for Mn 2$p$ core levels along with peak fitting is shown in Fig. 2b, where the open black circle are the experimental data the red line is overall envelop of the spectrum the solid blue lines are the Mn 2$p$ peaks are shown. The Mn 2$p$ spectrum shows two distinct peaks located at 642.2 eV and 654.1 eV corresponding to Mn 2$p$$_{3/2}$ and 2$p$$_{1/2}$ resulted from spin-orbital splitting of 2$p$ orbitals with splitting energy of 12.1 eV. The peak position reveals that the Mn cation present in +4 oxidation state and results are agreement with literature.\cite{cao}

Fig. 2c shows the core level spectra of Ho 4$d$ along with the fitting. In the figure open black circle is experimental data, solid red line is the overall envelop of the XPS spectrum and solid blue lines are the Ho 4$d$ orbitals. It is quite evident from the figure that the Ho 4$d$$_{5/2}$ peak is located at 160.6 eV. The detail analysis of XPS spectrum and peak postition of 4$d$ core level  reveals that the Ho cation are resent in +3 oxidation states and matches with earlier literature.\cite{ho}  The XPS study revails that the observed cationic oxidation state are Co$^{2+}$, Mn$^{4+}$ and Ho$^{3+}$ in Ho$_2$CoMnO$_6$.

\begin{figure}[th]
	\centering
		\includegraphics[width=8cm]{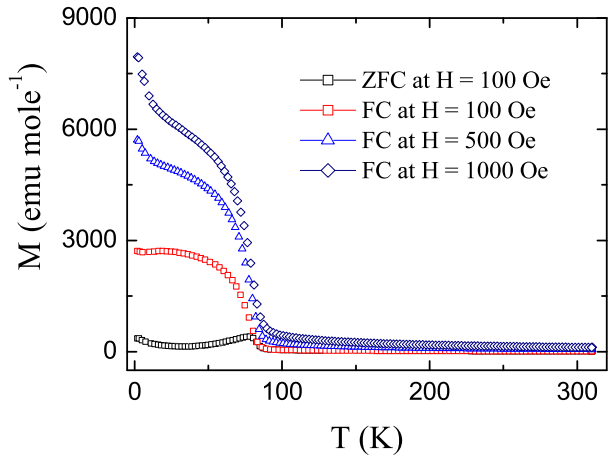}
\caption{(Color online) Temperature dependent magnetization data M(T) shown for Ho$_2$CoMnO$_6$ measured at different fields.}
	\label{fig:Fig3}
\end{figure}

\subsection{Magnetization study}

DC magnetization data was measured as both temperature dependent ($M(T)$) and field dependent at constant temperature ($M(H)$) for Ho$_2$CoMnO$_6$. Temperature variation of magnetization data is recorded in both zero field cooled ($ZFC$) and field cooled ($FC$) mode as shown in Fig. 3. In this study $ZFC$ data is measured in 300 K to 2 K range in an applied field of 100 Oe for Ho$_2$CoMnO$_6$ sample. $FC$ data is also measured at three different applied magnetic fields (see Fig. 3). The close observation of Fig. 3 shows that with decreasing temperature the magnetic moment ($M$) in $M(T)$ curve under both $ZFC$ and $FC$ mode is steady till $\sim$90 K, however with further lowering of temperature magnetic moment begin to rise below 90 K. The sharp rise in moment around below 90 K is marked by paramagnetic to ferromagnetic phase transition in Ho$_2$CoMnO$_6$. Further, it is worth to mention here that our XPS results reveals that Mn and Co cations are in +4 and +2 oxidation states respectively with outer electronic configuration Mn$^{4+}$ (t$_{2g}^{3}$e$_{g}^{0}$), Co$^{2+}$ (t$_{2g}^{5}$e$_{g}^{2}$). It is expected that these Mn$^{4+}$-Co$^{2+}$ will engage in superexchange interaction and would ferromagnetically ordered. It is quite evident from the figure that with decreasing temperature the $M_{ZFC}$ curve shows a peak like behavior at T$_c$, whereas $M_{FC}$ increases monotonically with lowering temperature. $M_{ZFC}$ and $M_{FC}$ curves measured at 100 Oe are overlapping on each other till 80 K, however with further decreasing temperature large bifurcation in $M_{ZFC}$ and $M_{FC}$ curves is observed. Further, $M_{ZFC}$ curve shows a little rise in magnetic moment at low temperature whereas $M_{FC}$  shows a slight down fall in moment at low temperature below 20 K which can be seen in $dM/dT$ vs $T$ plot (see inset Fig. 4). Since, the Ho$_2$CoMnO$_6$ shows a magnetic phase transition from typical like PM-FM with sharp rise in magnetic moment around 90 K. To precisely find the transition temperature we chose to plot the temperature derivative of magnetic moment against temperature i.e. $dM/dT$ vs $T$ plot shown in inset Fig. 4. The point of inflection or minima in the $dM/dT$ vs $T$ plot gives the transition temperature ($T_c$) = 80 for Ho$_2$CoMnO$_6$. Further, it is evident that the $dM/dT$ vs $T$ gives another local minima at low temperature which is often attributed to the ordering of moments on rare earth ions present in the material.\cite{sanch, kaka} We have measured $M_{FC}$ data at different applied field we observe that with increasing field T$_c$ shifts to higher field which can be understood assuming that the forced spin arrangement happens in field direction. It is further worth considering that at low temperature the feature of $M_{FC}$ changes with increasing field. It is evident in Fig. 3 $M_{FC}$ measured at 500 Oe and 1000 Oe shows a sharp increase in magnetic moment below 20 K, which at low field shows a downfall in moment. The possible reason of this increase in moment at higher field may be due to spin flipping of rare earth ions and favoring ferromagnetic ordering with Mn/Co sublattice.

\begin{figure}[t]
	\centering
		\includegraphics[width=8cm]{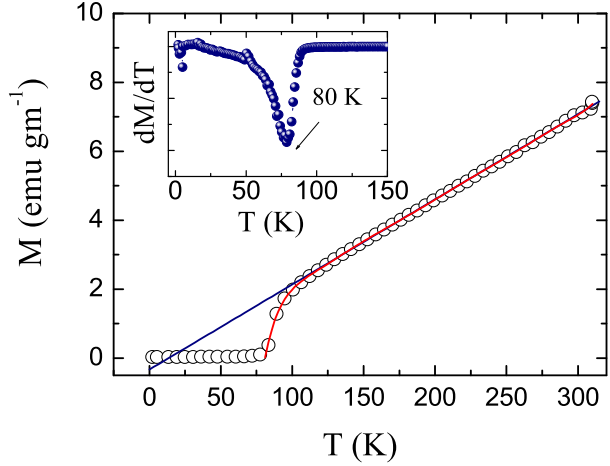}
\caption{(Color online) M(T) data plotted in terms of inverse susceptibility $\chi^{-1}$, black and red solid lines are fitting due to   Curie Weiss Law and modified Curie Weiss Law respectively. Inset shows dM/dT vs T plot showing T$_c$ for Ho$_2$CoMnO$_6$.}
\label{fig:Fig4}
\end{figure}

\begin{figure}[t]
	\centering
		\includegraphics[width=8cm]{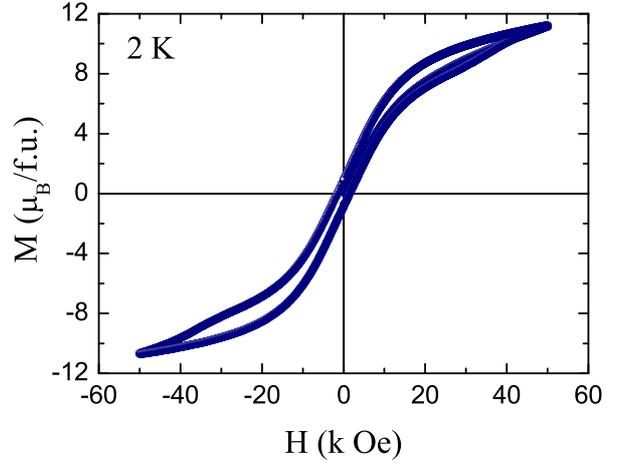}
\caption{(Color online) Isothermal data collected at 2 K in applied field of $\pm$ 50 kOe for Ho$_2$CoMnO$_6$.}
	\label{fig:Fig5}
\end{figure}

\begin{figure*}[th]
	\centering
		\includegraphics[width=14cm]{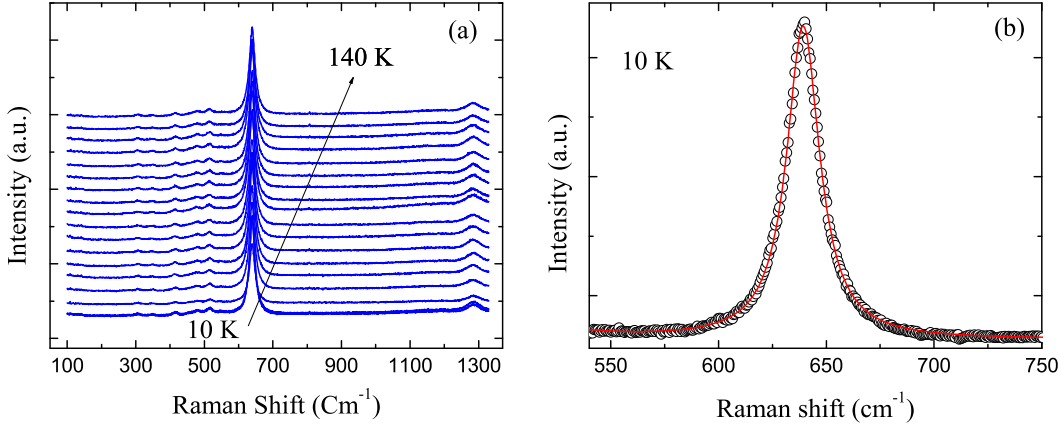}
\caption{(Color online) (a) Raman spectra of Ho$_2$CoMnO$_6$ measured at different temperatures.   (b) shows the line shape and its Lorentzian fitting of A$_{1g}$ and B$_{2g}$ Raman modes at 494 and 644 cm$^{-1}$ respectively.}
	\label{fig:Fig6}
\end{figure*}

\begin{figure}[th]
	\centering
		\includegraphics[width=8cm]{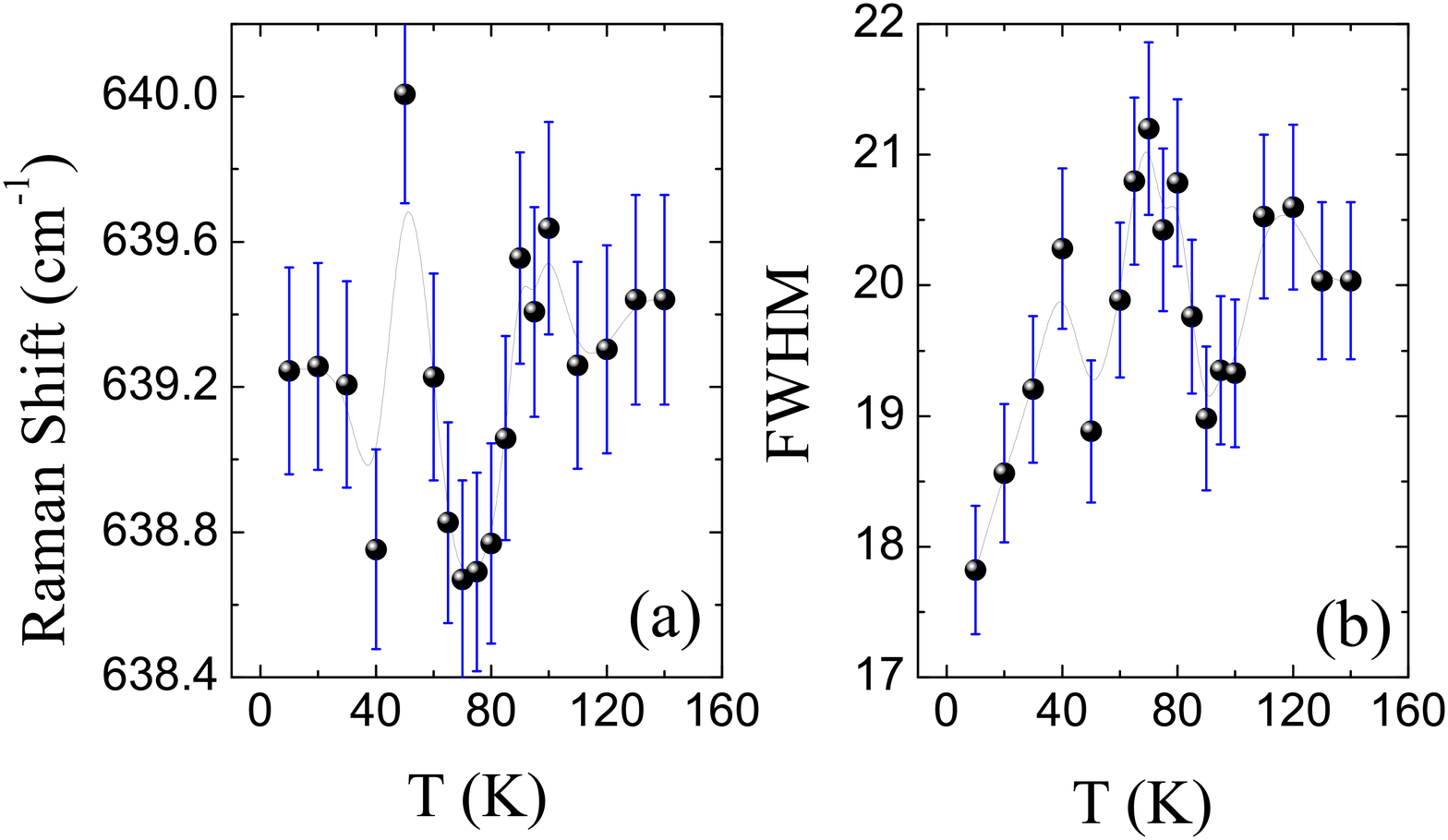}
\caption{(Color online) Tempeature Variation of (a) Raman shift (b) FWHM for Raman mode at 640 cm$^{-1}$ corresponding to stretching of Co/MnO$_6$ for Ho$_2$CoMnO$_6$. The solid line is fitting due to Eq. 2}
	\label{fig:Fig7}
\end{figure}

\begin{figure*}[th]
	\centering
		\includegraphics[width=14cm]{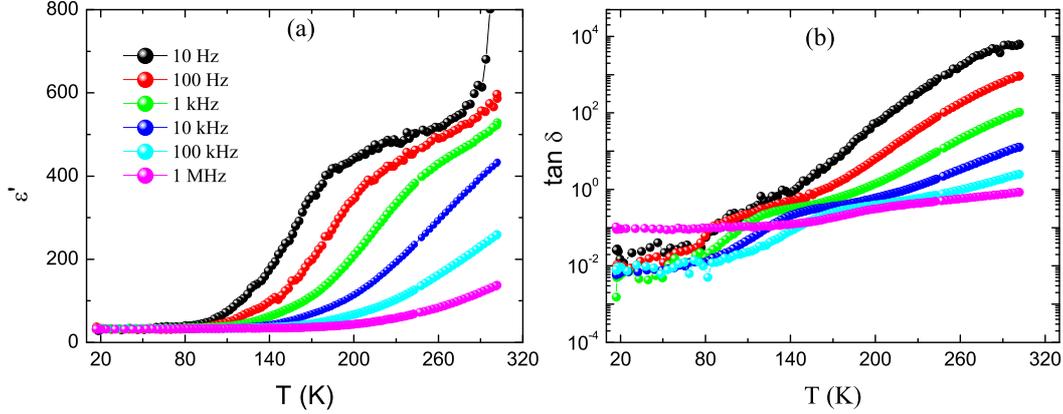}
\caption{(Color online) Temperature dependent (a) real part of complex dielectric permittivity ($\epsilon$$^\prime$) (b) loss tangent (tan $\delta$) measure for Ho$_2$CoMnO$_6$ in the temperature range of 20 K to 300 K at various frequencies.}
	\label{fig:Fig8}
\end{figure*}

\begin{figure}[h!]
	\centering
		\includegraphics[width=8cm]{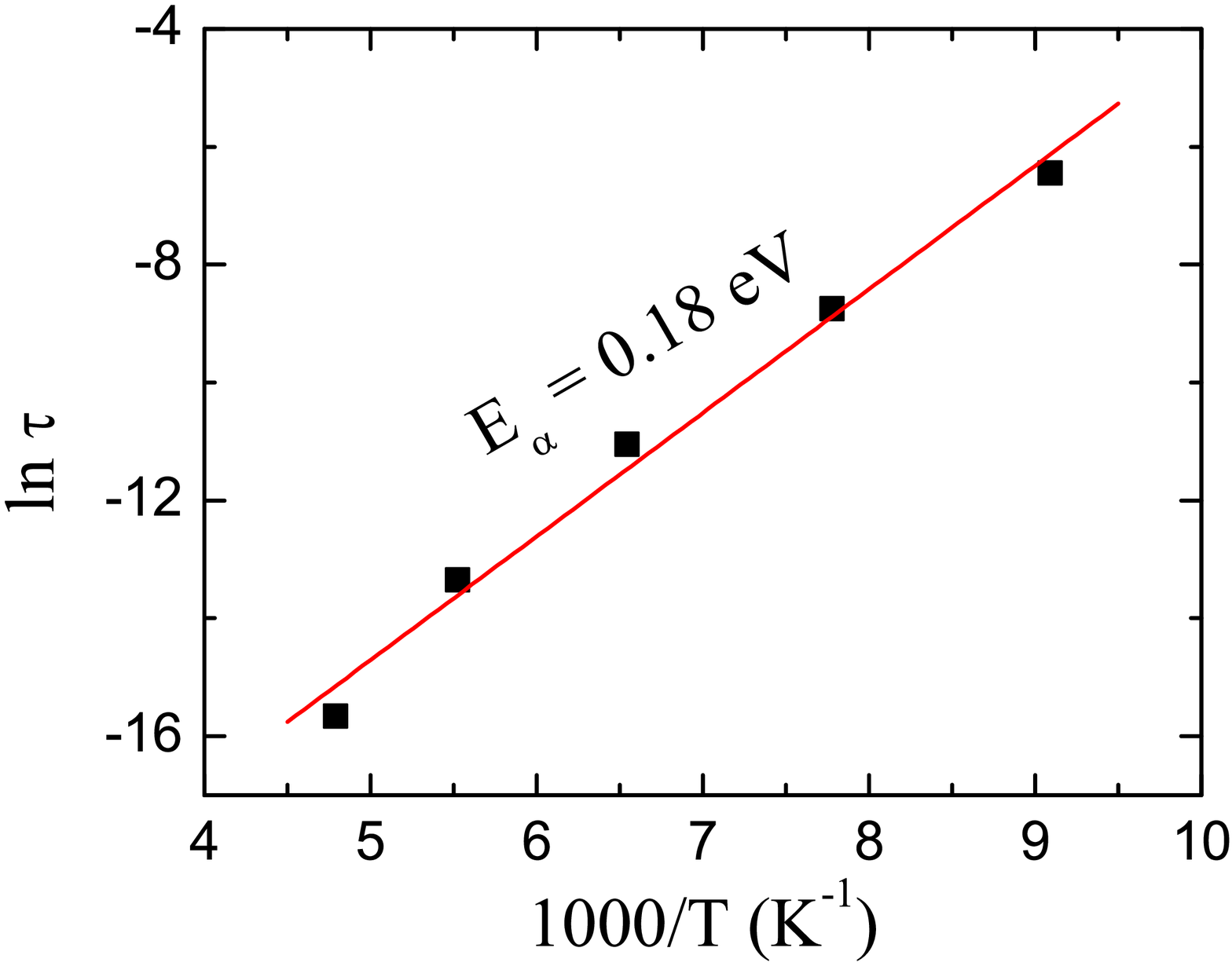}
\caption{(Color online) Variation of relaxation time against normalized temperature i.e ln $\tau$ vs 1000/T obtained from tangent loss plot.}
	\label{fig:Fig9}
\end{figure}

To understand magnetic behavior in further depth we have plotted magnetization data in terms of temperature dependent inverse magnetic susceptibility i.e. $\chi^{^{-1}}$ vs $T$ as shown in Fig. 4. In paramagnetic region close above T$_c$ we see inverse magnetic susceptibility data shows deviation from linearity unlike expected for paramagnetic system by Curie Weiss (CW) law.  Such failure from linearity of inverse susceptibility in paramagnetic state have also been reported for Ho$_2$NiMnO$_6$.\cite{retu} In paramagnetic region above T$_c$ it is generally believe that the double perovskite compound with rare earth ions at A-site shows such deviation in inverse magnetic susceptibility. The conventional Curie Weiss law fails to described this behavior successfully (see the straight line fitting using conventional CW law represented by black solid line in Fig. 4) thus a modified CW law is employed to analyze the magnetic susceptibility:\cite{booth}

\begin{eqnarray}
\chi = \frac{C_{TM}}{T - \theta_{TM}} + \frac{C_{RE}}{T - \theta_{RE}}
\end{eqnarray}

Here $C_{TM}$ and $\theta_{TM}$  is Curie Constant and paramagnetic Curie temperature for transition metal sublattice. Where $C_{RE}$ and $\theta_{RE}$ are the Curie Constant and paramagnetic Curie temperatures for rare earth sublattice. We have performed the non-linear fitting of susceptibility data  measured at 100 Oe for  Ho$_2$CoMnO$_6$. The fitting is done above $T_c$ in the temperature range 80 K to 300 K. The fitting parameters obtained from fitting of $\chi^{-1}$ with Eq. 1 in Fig. 4 (see solid red line) gives the value of $C_{TM}$ and $\theta_{TM}$ as 3.452(3) emu K mole$^{-1}$ Oe$^{-1}$ and 80.06 K respectively. The effective paramagnetic moment ($\mu_{eff}^{TM}$) is calculated using these fitting parameters come out to be 5.202(5) $\mu_B/f.u.$, the obtained values are in close to theoretical calculations.  Further, we have observed that the values of $C_{RE}$ and  $\theta_{RE}$ for Ho$_2$CoMnO$_6$ are 38.87 emu K mole${_1}$ Oe$^{-1}$ and -8.2(1) K respectively. The negative value of $\theta_RE$ signifies that the Ho$^{3+}$ spins are antiferromagnetic ordering relative to Mn/Co sublattice. However, the effective magnetic moments for rare earth ($\mu_{eff}^{RE}$) is 17.4 $\mu_B/f.u.$ which is higher then the effective magnetic moment for free Ho$^{3+}$ ions. The modified Curie Weiss law in Eq. 1 is presented considering non-interacting Mn/Co and Ho sublattics in this compound. However, the higher value of $\mu_{eff}^{RE}$ possibly due to interaction of Mn/Co and Ho ions at low temperature. However, the detail understanding of this fact require investigation using local probing techniques. 

Fig. 5 shows the isothermal magnetization $M(H)$ data i.e. $M vs H$ plot collected at 2 K up to $\pm$50 kOe applied magnetic field.  Isothermal magnetization curve shows hysteresis which is signature of ferromagnetic ordering. However the $M(H)$ curve is asymmetric about origin. Further the magnetic moment does not show any signature of saturation even at highest applied magnetic field of 50 kOe. The magnetic moment at 50 kOe is about 11.16(5) $\mu$$_B$/f.u. where as the remanent magnetization and coercive force is 1.009 $\mu$$_B$/f.u. and 2 kOe respectively.

\subsection{Temperature dependent Raman study}
It is well know fact that most materials with magnetic ordering shows to spin phonon coupling.\cite{ilive, sing, kumar, liu} The spin phonon coupling can be understood from Raman study as the deviation of mode position and  the line width from anharmonic behavior.\cite{ilive} The change in Raman mode frequency, line width and modulation in intensity of a particular mode is associated with incommensurate modulation of the atomic positions as explained by  Slawinski \textit{et al}.\cite{slw} In many ferromagnetic compounds with perovskite/double perovskite structure strong spin phonon coupling have been reported earlier.\cite{kumar, liu, laver} Ho$_2$CoMnO$_6$ shows  ferromagnetic magnetic ordering around 80 K as discussed in magnetization section above. The sharp magnetic transition is associated with ordering of Co/Mn spins by superexchange interaction via Co$^{2+}$-O-Mn$^{4+}$. This motivated us to carry out the Raman study to investigate the presence of spin phonon coupling in this material. 

\begin{figure*}[h!]
	\centering
		\includegraphics[width=14cm]{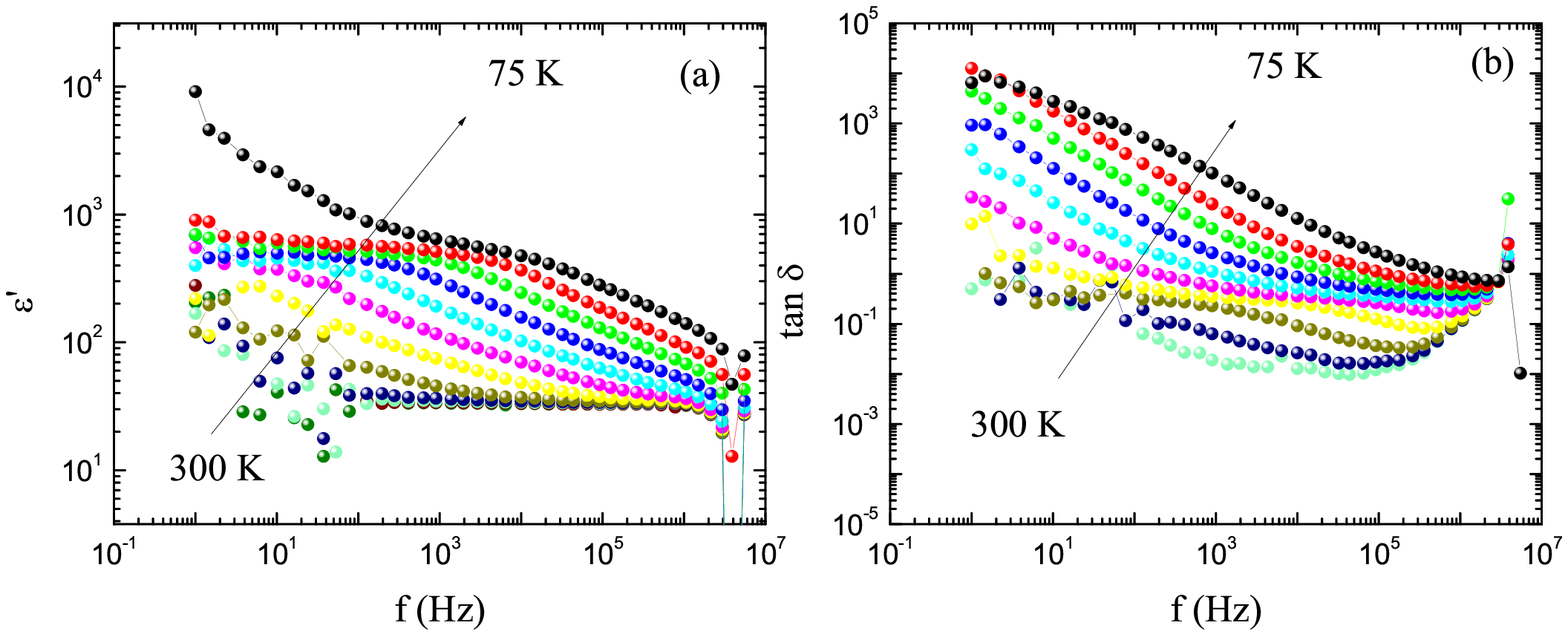}
\caption{(Color online) Frequency dependent (a) real part of complex dielectric permittivity ($\epsilon$$^\prime$) (b) loss tangent (tan $\delta$) measure for Ho$_2$CoMnO$_6$ at various temperatures between 25 K and 300 K in the frequency range 1 Hz to 5.6 MHz.}
	\label{fig:Fig10}
\end{figure*}

Fig. 6a shows the temperature dependent Raman spectra collected at various temperatures, including magnetic transition temperature $T_c$ around 80 K for Ho$_2$CoMnO$_6$. The Raman spectra shows a intense phonon mode at 640 cm$^{-1}$ which is focus of our investigation in this study. This Raman mode is known as  A$_{1g}$ breathing mode, associated with the symmetric stretching of the (Co/Mn)O$_6$ octahedra.\cite{ilive} Additionally, the other weak modes at 494 cm$^{-1}$ and  1278 cm$^{-1}$ are due to bending/ rotation of Co/MnO$_6$ octahedra and second-order overtones of the breathing mode respectively.\cite{meyer} The temperature dependent Raman spectra shows a evolution of phonon modes and line width. We have analyzed the raman spectra at all temperatures by fitting with Lorentzian function. The peak shape for A$_{1g}$ Raman mode along with Lorentzian fitting is shown in Fig. 6b for Raman spectrum at 100 K temperature. From the fitting we have obtained the temperature evolution of Raman mode frequency and line width. Fig. 6a and 6b shows the temperature evolution of phonon mode position and line width for A$_{1g}$ breathing mode for Ho$_2$CoMnO$_6$. It is well exhibited in the Fig. 7 that both the phonon mode position and line width shows the deviation at magnetic ordering temperature. Such deviation is due to spin phonon coupling in ferromagnetic ordered systems. 

\begin{figure}[th]
	\centering
		\includegraphics[width=8cm, height=12cm]{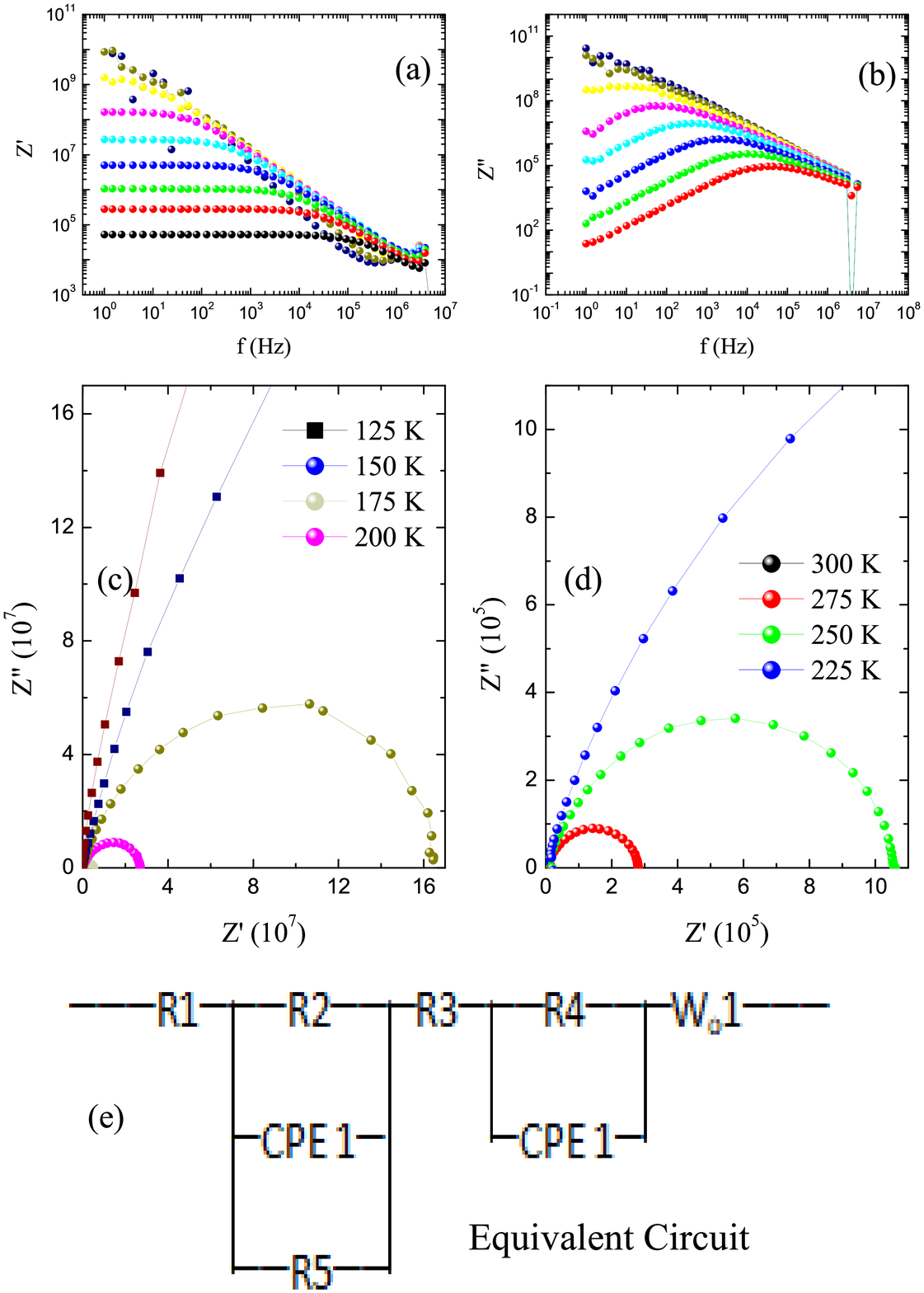}
\caption{(Color online) (a) Frequency dependent real part of impedance $Z^{\prime}$ measure at different temperatures. (b) Frequency dependent imaginary part of impedance $Z^{\prime\prime}$ measure at various temperatures. (c), (d) real $Z^{\prime}$ and imaginary part $Z^{\prime\prime}$ plotted in terms of Nyquist plot $Z^{\prime}$ vs $Z^{\prime\prime}$. (e) Equivalent circuit of Nyquist plot for Ho$_2$CoMnO$_6$}
	\label{fig:Fig11}
\end{figure}
\subsection{Dielectric study}
To understand the dielectric response of the nano-crystalline Ho$_2$CoMnO$_6$  we have carried out the detail measurements. The temperature dependent real part of complex dielectric permittivity $\epsilon^{\prime}$ and tan $\delta$ presented in  Fig. 8a and 8b respectively. The measurements are performed in the temperature range of 20 K to 300 K at different frequencies for Ho$_2$CoMnO$_6$. Further, for relaxor systems with relaxation mechanisms, each relaxation component will correspond to plateaus in $\epsilon^{\prime}$(T) and respond with peaks in tangent loss factor (tan$\delta$). For this material we have observed that with increasing temperature the $\epsilon^{\prime}$ at low temperatures  $\epsilon^{\prime}$ increases slowly. However with increasing temperature $\epsilon^{\prime}$ increases sharply. Further, with increasing frequencies $\epsilon^{\prime}$ decreases sharply. The higher value of $\epsilon^{\prime}$ at low frequency is attributed to the accumulation of the charges at grain boundaries  On careful observation of tangent loss curve tan $\delta$ it is clearly seen that there is a broad hump at low temperature which is feature of relaxor phenomenon. The observed relaxation is frequency dependent and shift to higher temperature with increasing frequency. To understand the mechanism of relaxation process we have picked the temperature corresponding to the peak value in the tan$\delta$ and plotted against the frequency/relaxation time. The relaxation time $\tau$ can be calculated from frequency at which tan$\delta$ peak is observed using the formula $\tau$ = 1/2$\pi$f. The relaxation time $\tau$ is plotted as a function of absolute temperature shown in Fig. 9. The relaxation time is distributive i.e. the peak in tangent loss is shifted to higher temperature with increasing frequency. This fact indicates that the relaxation mechanism is thermally activated.

The relaxation mechanisms and its origin can be analyzed by fitting the data presented in Fig. 9 with Arrhenius law given by, $\tau_{tan \delta}$ = $\tau_0$ exp(-$E_a$/$K_{B}$$T$) where, $T$ is the temperature where peak occurs in tangent loss curve at a particular frequency $f_{tan \delta}$, $\tau_0$ and $E_a$ are characteristic frequency and activation energy respectively and, $k_B$ is the Boltzmann constant. From the fitting parameters obtained in Fig. 9 we have calculated the activation energy E$_a$ = 0.18 eV.

To further understand the dielectric response we have measure the frequency dependent $\epsilon$$\prime$ and tan $\delta$ over the frequency range 1 Hz to 5.5 MHz for Ho$_2$CoMnO$_6$ at different temperatures ranging from 50 K to 300 K. In the Fig. 10a we observed that Ho$_2$CoMnO$_6$ exhibits high dielectric constant at low frequency and at high temperature. The dielectric spectrum shown in Fig. 10a clearly shows strong dispersion indicated at low frequency which moves to higher frequency with decreasing temperature. The separate plateau in Fig. 10a are attributed to static and optical dielectric constant. The dielectric loss have be shown in Fig. 10b its is found that the dielectric loss shows similar response.

\subsection{Impedance spectroscopy}

Fig. 11a shows the real part of complex impedance ($Z^{\prime}$) plotted as a function of frequency in the frequency range 1 Hz to 5.5 MHz at various temperatures between 50 K to 300 K. For clarity both the axis are in logarithmic scales. Impedance spectroscopy can be utilized to distinguish the contributions to the dielectric properties from  grain, grain boundaries and electrode-sample contact interface. It is quite evident from the figure that the $Z^{\prime}$ decreases with increasing temperature. At low temperature $Z^{\prime}$ remains almost constant with increasing frequency, however with further increase in frequency it decreases sharply. With increasing temperature the constant region in the $Z^{\prime}$ i.e frequency independent region decreases. It is observed that, the frequency independent region moves to higher frequency with increasing temperature. Further, it is observed that at higher frequency and high temperature the $Z^{\prime}$ almost merg into one set of plot. This feature is possibly due to the release of accumulated space charges at high temperatures hence contribute to the enhancement of conduction in this material at high temperature. Imaginary part of impedance ($Z^{\prime\prime}$) is shown in Fig. 11b for wide frequency range. $Z^{\prime\prime}$ shows a peak which attain $Z^{\prime\prime}_{max}$ for all the curves measured at different temperatures, further it is evident that the peak moves towards higher frequency with increasing temperature. The peak shift to higher frequency with increasing temperature suggests that the relaxation time constant decreases with increasing temperature. The relaxation time is distributive and suggests the thermally activated relaxation mechanism.

\begin{figure}[h!]
	\centering
		\includegraphics[width=8cm]{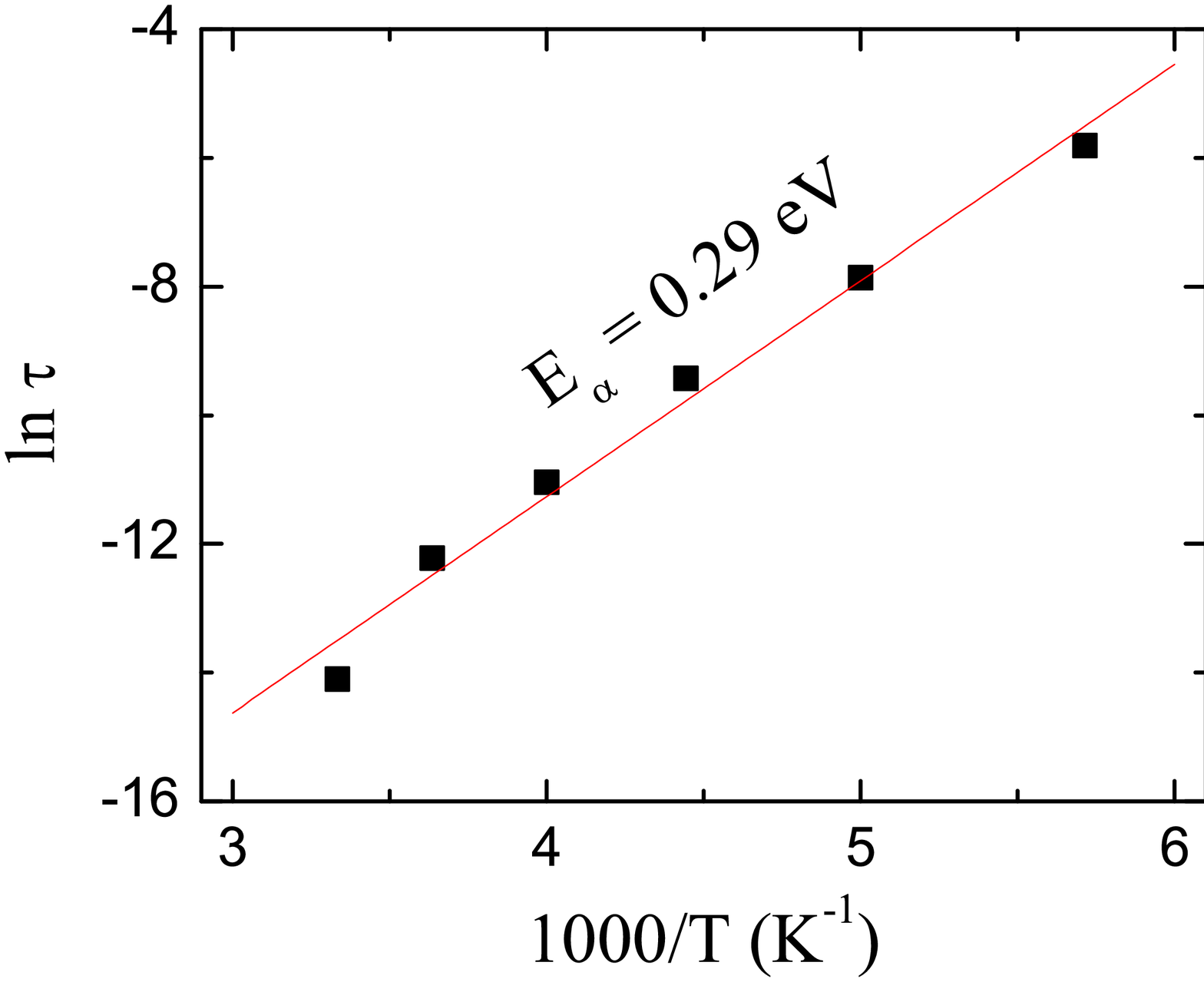}
\caption{(Color online) Variation of relaxation time against normalized temperature i.e ln $\tau$ vs 1000/T obtained from Z$^\prime\prime$ plot.}
	\label{fig:Fig12}
\end{figure}

We know that the most probable relaxation time ($\tau$) can be determined for relaxation system by identification the position of the loss peak in the $Z^{\prime\prime}$ vs log ($f$) plots using the relation:

\begin{eqnarray}
\tau = \frac{1}{\omega} = \frac{1}{2 \pi f}
\end{eqnarray}

where $\tau$ is relaxation time and $f$ is the relaxation frequency. 
To further understand the relaxation behavior we have plotted the relaxation time $\tau$ vs temperature 10$^3$/$T$ (K$^{-1}$). Fig. 12 shows the $\tau$ plotted as a function of absolute temperature obtained from imaginary part of complex impedance. It is observed that the relaxation time follows Arrhenius behavior given as:
 
\begin{eqnarray}
\tau_{b} = \tau_{0}] exp{\left( \frac{-E_a}{k_BT} \right)}
\end{eqnarray}

where $\tau_0$ is the pre-exponential factor, k$_B$ the Boltzmann constant and $T$ the absolute temperature. From the fitting parameters the activation energy (E$_a$) have been calculated and is found to be 0.29 eV.

Fig. 11c and 11d shows the $Z^{\prime}$ vs $Z^{\prime\prime}$ in the form of Nyquist plots at some selective temperatures measure in the wide frequency range 1 Hz to 5.5 MHz. It is quite evident from the figures that the plot gives the semicircle in whole range of temperature. It is found that the radius of Nyquist plot decreases with increasing temperature. The semicircle in the Nyquist plots have the origin on the x-axis which suggest that the Debye type of relaxation is found in this material. It also manifests that there is a distribution of relaxation time instead of a single relaxation time in the material. The experimental results reveals that the material is deviated from ideal Debye's behavior. The Equivalent circuit for the Nyquist plot represented in Fig. 11c and 11d  is given in Fig. 11(e). The equivalent circuit consists of series and parallel combination of resistor, constant phase change and Warburg Open. 

\begin{figure*}[th]
	\centering
		\includegraphics[width=14cm]{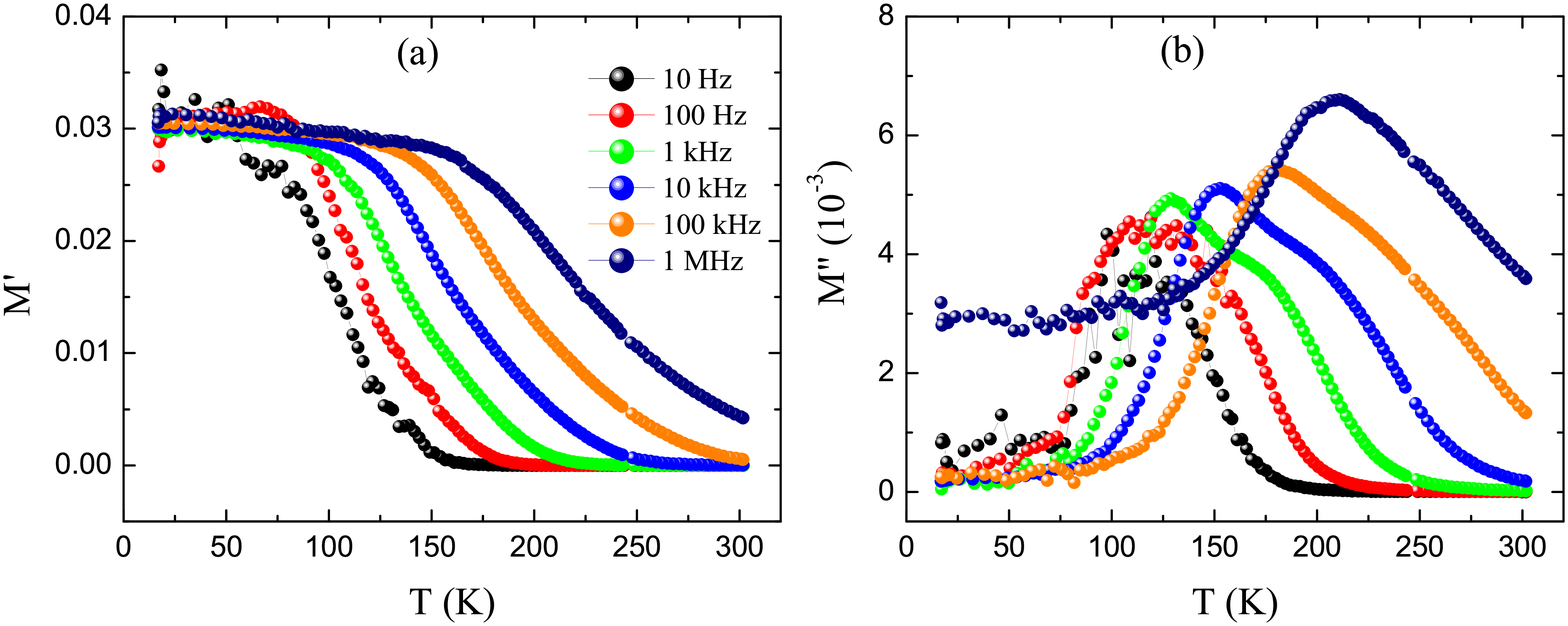}
\caption{(Color online) (a) Variation of real part of electrical modulus ($M^\prime$) with temperature. (b) Imaginary part of electrical modulus $M^{\prime\prime}$ as a function of temperature.}
	\label{fig:Fig13}
\end{figure*}

\begin{figure}[th]
	\centering
		\includegraphics[width=8cm]{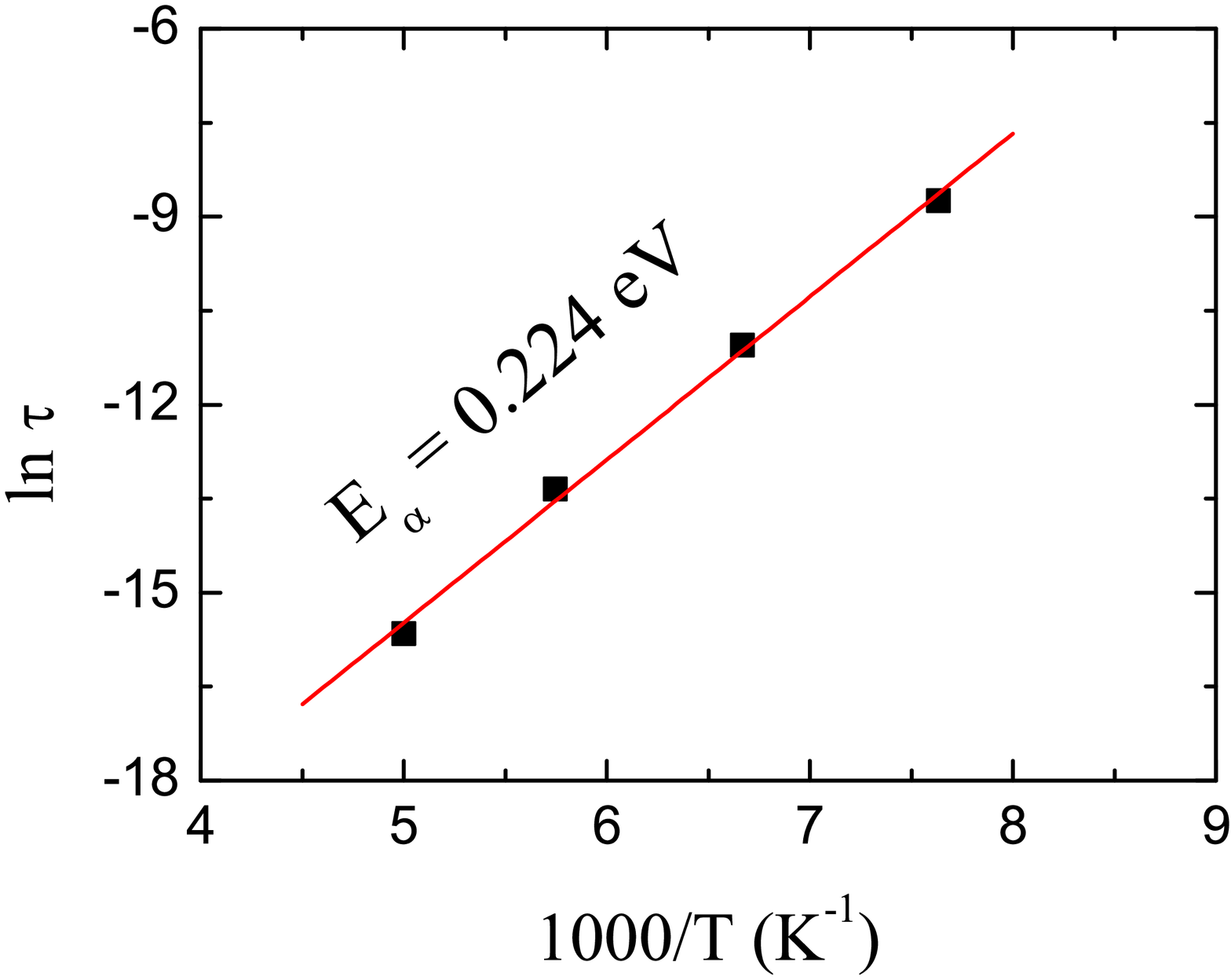}
\caption{(Color online) Variation of relaxation time against normalized temperature i.e ln $\tau$ vs 1000/T obtained from M$^\prime\prime$ plot.}
	\label{fig:Fig14}
\end{figure}

\begin{figure}[th]
	\centering
		\includegraphics[width=8cm]{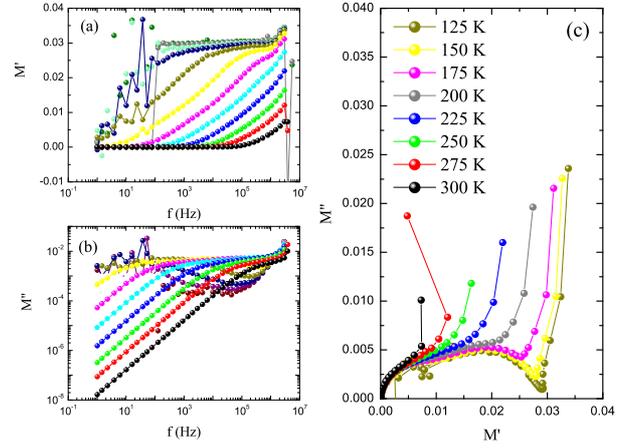}
\caption{(Color online) (a,b) shows real and imaginary parts of electrical modulus as a function of frequency. (c) shows the niqust plot of electrical modelus i.e. M$^\prime$ vs M$^\prime\prime$.}
	\label{fig:Fig15}
\end{figure}

\subsection{Electrical modulus}
Information of interface polarization, relaxation time, electrical conductivity and grain boundary
conduction effects etc can be deduced from  electrical modulus of materials. Figs. 13a and 13b show the temperature dependent real ($M^{\prime}$) and imaginary ($M^{\prime\prime}$) part of electrical modulus obtained at selective frequencies for Ho$_2$CoMnO$_6$ in the temperature range of 20 K to 300 K. Its evident from the Fig. 13a that the real part of electrical modulus is steady at low temperature for all frequencies. It further found that with increasing temperature the  $M^{\prime}$ decreases sharply. with increasing frequency the steady feature at low temperature moves to higher temperature. There is large dispersion in the $M^{\prime}$ curves at temperature above 100 K. At 300 K the $M^{\prime}$ is further merges into one curve. Further, Fig. 13b shows the variation of $M^{\prime\prime}$ as function of temperature. From the figure it is evident that $M^{\prime\prime}$ shows a peak position in the  $M^{\prime\prime}$ vs T plot. It is further observed that the peak position in $M^{\prime\prime}$ data moves to the higher temperature with increasing frequency.  Moreover, The $M^{\prime\prime}_{max}$ increases with increasing frequency. The shifting of peak to higher temperature with increasing frequency shows that the relaxation mechanism in this material is thermally activated. The frequency and corresponding temperature at which peak position in $M^{\prime\prime}$ vs T data is found is piked up and plotted in in Fig. 14. The relaxation time $\tau$ can be obtained from frequency $f$ using formula $\tau$ = 1/2$\pi$$f$. Fig. 14 shows the variation of relaxation time as function of absolute temperature i.e. $\tau$ vs 1000/T. It is found that the relaxation time obeys and fitted well with Arrhenius law given as:

\begin{eqnarray}
\tau vs \tau_0exp\left(-E_a/k_BT\right) 
\end{eqnarray}

where $\tau$ is relaxation time, E$_a$ is activation energy k$_B$ is the Boltzmann constant.  The relaxation time obtained from the $M^{\prime\prime}$ is well fitted with the Arrhenius law for thermally activated mechanisms. The activation energy is calculated from the fitting parameters and we found that the E$_a$ = 0.29 eV in agreement of with our impedance spectroscopy results.
 
Fig. 15 summarize the variation of $M^{\prime}$ and $M^{\prime\prime}$ with frequency at selected temperatures and plot of $M^{\prime}$ vs $M^{\prime\prime}$ on a complex plane. Once again, $M^{\prime\prime}$ spectroscopy plot reveals relaxation phenomena in the material. The maximum value ($M^{\prime\prime}$) of the $M^{\prime\prime}$ peak shifts to higher frequency, which suggests that hopping of charge carriers is predominantly thermally activated. Asymmetric broadening of the peak indicates spread of relaxation with different time constants, which once again suggests the material is non-Debye-type.
Fig. 15c shows the Nyquist plot for electrical modulus. It si observed that the relaxation time is distributed and thus the feature suggest that the materals non-Debye's in nature.
  
\begin{figure}[th]
	\centering
		\includegraphics[width=8cm]{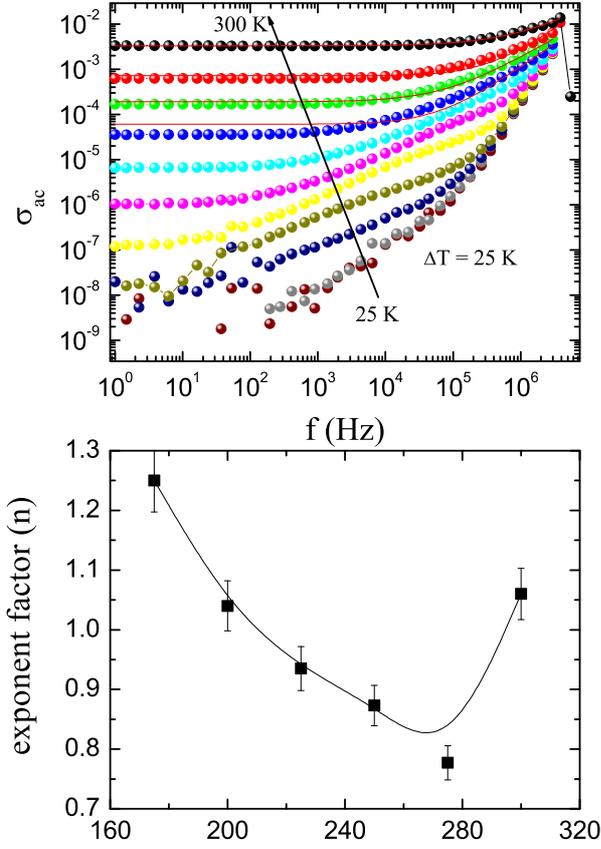}
\caption{(Color online) Frequency dependence plot of the  conductivity ($\sigma_{ac}$) for temperatures ranging from 50 K to 300 K are shown for Ho$_2$CoMnO$_6$. solid red lines are fitting due to Eq. 6. (b) Figure shows the exponent $n$ versus the temperature for nano-crystalline Ho$_2$CoMnO$_6$}
	\label{fig:Fig16}
\end{figure}

\begin{figure}[th]
	\centering
		\includegraphics[width=8cm]{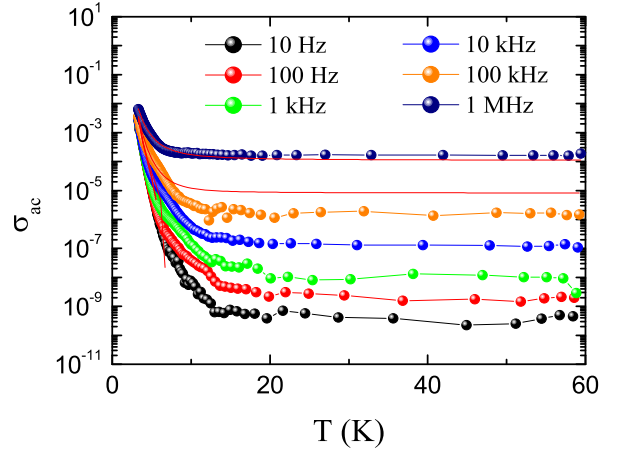}
\caption{(Color online) The variation of $\sigma_{ac}$ with absolute temperature (10$^3$/T) is shown for Ho$_2$CoMnO$_6$.}
	\label{fig:Fig17}
\end{figure}

\subsection{Electric conductivity}
To further understand the charge hoping and electrical properties we have investigated the AC conductivity in Ho$_2$CoMnO$_6$. The AC conductivity is calculated by using relation $\sigma_{ac} = \epsilon_0 \omega \epsilon^{\prime\prime}$.\cite{sing} Fig. 16a shows the variation of AC conductivity with frequency i.e. $\sigma_{ac}$ vs $f$ at some selective temperatures in the range 50 K to 300 K. It is evident from the figure that at low frequencies the conductivity is independent of frequency and gives a plateau region at all temperatures. In this region of frequency the conduction is mainly dominated by DC conductivity ($\sigma_{dc}$). However, at higher frequencies the conductivity increases with increasing frequency. It is further notable that the  plateau region marked by DC conductivity in the Fig. 16a extends to the higher frequencies with increasing temperature. The frequency independent region also suggests that the hopping charges carrier is absent at low frequencies. The AC conductivity at high frequency in this case obey Jonscher‟s Universal Power Law given as follow:\cite{thakur}

\begin{eqnarray}
\sigma_{ac} = \sigma_{dc} + A\omega^n
\end{eqnarray}

when A is a temperature dependent constant, $\omega$ = 2$\pi$$f$ and $n$ is the power law exponent which generally varies between 0 and 1 depending upon temperature. The value of power law exponent $n$ represent the extent of interaction between mobile ions and lattice around. For non-interacting Debye system $n$ = 1 and with decreasing $n$ value the interaction is expected to increase between lattice surrounding and mobile ions. Further the constant A  gives the degree of polarizibility. 

The frequency dependent $\sigma_{ac}$  is fitted with the power law as shown in Fig. 16a. The conductivity data is fitted well  in full frequency range 1 Hz to 5.5 MHz. From the fitting parameter we have calculated the exponent $n$. The Fig. 16b shows the variation of $n$ with temperature its is observed that the $n$ is less then 1 at high temperature and with decreasing temperature it increases. The variation of $n$ with temperature is suggestive of non-Debye's type behavior.

The variation of AC conductivity with inverse absolute temperature i.e.  $\sigma_{ac}$ vs 10$^3$/T at some selective frequencies is shown in Fig. 17. We observed that with increasing frequency the conductivity increases. the conductivity data is fitted with following equation:

\begin{eqnarray}
\sigma_{ac} = \sigma_{0}exp\left(\frac{-E_a}{k_BT}\right)
\end{eqnarray}

where $\sigma_{0}$ is pre-exponent factor, $k_B$ is Boltzmann constant and $E_a$ is the activation energy.
It is found that the temperature dependent data is fitted at high temperatures for all frequencies as shown by solid lines in Fig. 17. From the fittng parameters we have calculated the activation energy for conductivity measured at different frequencies and are found that with increasing frequency the activation  decreases. The obtained activaton energies are listed here $E_a$(1 MHz) = 1.19 eV, $E_a$(100 kHz) = 2.90 eV, $E_a$(10 kHz) = 3.54 eV, $E_a$(1 kHz) = 3.60 eV, $E_a$(100 Hz) = 3.68eV and $E_a$(10 Hz) = 3.70 eV.

\section{Conclusion}
Ho$_2$CoMnO$_6$ nano-crystalline was successfully synthesised by sol-gel method. Structural, magnetic and dielectric properties were studied.  Structural analysis of XRD shows the samples crystallize in monoclinic structure with \textit{P2$_1$/n} space group with average crystallite size for present compound is $\sim$76.65 nm. Charge state of various elements present in the material is confirmed using X-ray photoelectron spectroscopy. Magnetization study shows that the material is ferromagnetic in nature and shows phase transition around 85 K. The effective magnetic moment obtained experimentally as 5.25 $\mu_B$/f.u. is found close to calculated value. Raman study shows spin phonon coupling is present as indicated by mode frequency deviation from anharmonic behavior across $T_c$. Dielectric measurements have been performed on Ho$_2$CoMnO$_6$ both temperature and frequency dependent. We found a very high dielectric constant which increases with decreasing frequency. The tangent loss shows the relaxation mechanism is active and it follows thermally activated behavior. The impedance spectroscopy reveals that the relaxation time is distributive and shows deviation from ideal Debye model. AC conductivity have been studied as a function of both temperature and frequency. We found that the conduction mechanism is obeyed by Jonscher's model. The exponent factor $n$ suggest that the material deviates from the ideal Debye model.

\section{Acknowledgment}
We acknowledge MNIT Jaipur, India for XPS data, AIRF (JNU) for magnetic measurement and SEM facilities. We acknowledge UGC-DAE-Consortium Indore and Dr. V. G. Sathe for Raman data. We also acknowledge Dr. A. K. Pramanik for dielectric measurement and UPEA-II funding for LCR meter. Author Ilyas Noor Bhatti acknowledge University Grants Commission, India for financial support.

\end{document}